\documentclass[fleqn,10pt,twocolumn]{wlscirep}
\usepackage[utf8]{inputenc}
\usepackage[T1]{fontenc}
\usepackage{braket}
\usepackage{amsbsy}
\title{Giant elastocaloric effect at low temperatures in TmVO$_4$ and implications for cryogenic cooling }

\author[1,2,*]{Mark P. Zic}
\author[1,3]{Matthias S. Ikeda}
\author[1,3]{Pierre Massat}
\author[4]{Patrick M. Hollister}
\author[1,3]{Linda Ye}
\author[1,3]{Elliott W. Rosenberg}
\author[1,3]{Joshua A. W. Straquadine}
\author[4,5]{Brad J. Ramshaw}
\author[1,3,+]{Ian R. Fisher}

\affil[1]{Geballe Laboratory for Advanced Materials, Stanford University, Stanford, CA 94305}
\affil[2]{Department of Physics, Stanford University, Stanford, CA 94305}
\affil[3]{Department of Applied Physics, Stanford University, Stanford, CA 94305}
\affil[4]{Laboratory of Atomic and Solid State Physics, Cornell University, Ithaca, NY 14853}
\affil[5]{Canadian Institute for Advanced Research, Toronto, Ontario M5G 1Z8, Canada}

\affil[*]{zic@stanford.edu}

\affil[+]{irfisher@stanford.edu}

\begin{abstract}
Adiabatic decompression of para-quadrupolar materials has significant potential as a cryogenic cooling technology. We focus on TmVO$_4$, an archetypal material that undergoes a continuous phase transition to a ferroquadrupole-ordered state at 2.15 K. Above the phase transition, each Tm ion contributes an entropy of $k_B \ln{2}$ due to the degeneracy of the crystal electric field groundstate. Owing to the large magnetoelastic coupling, which is a prerequisite for a material to undergo a phase transition via the cooperative Jahn-Teller effect, this level splitting, and hence the entropy, can be readily tuned by externally-induced strain. Using a dynamic technique in which the strain is rapidly oscillated, we measure the adiabatic elastocaloric coefficient of single-crystal TmVO$_4$, and thus experimentally obtain the entropy landscape as a function of strain and temperature. The measurement confirms the suitability of this class of materials for cryogenic cooling applications, and provides insight to the dynamic quadrupole strain susceptibility.
\end{abstract}
\begin{document}

\flushbottom
\maketitle
\thispagestyle{empty}

Interest in elastocaloric cooling has primarily focused on refrigeration and air conditioning near room temperature, and has typically utilized the latent heat associated with first-order phase transitions \cite{Chen2021-shapeMemoryReview}. However, there is also considerable potential to employ elastocaloric effects for cooling at much lower, cryogenic temperatures. Low temperature cooling of materials using pressure and strain has been explored previously within the context of materials that couple to these quantities either partially or indirectly\cite{Strassle2000-CeSbECE,Strassle2002-Ce3Pd20Ge6ECE,Strassle2003-MCEviaPressure,Muller1998-PrLaNiO3ECE,Shepherd1965-ECEofOHmolecules}. Here, we explore the potential of adiabatic decompression of Jahn-Teller active materials, for which strain couples directly (bilinearly) to the active quadrupolar degrees of freedom, for cryogenic cooling. The requisite large elastocaloric effect (ECE) in this case is provided by the presence of a large quadrupole strain susceptibility, which is implicit in such materials in the temperature regime above any associated phase transitions. The cooling effect is analogous to that achieved by adiabatic demagnetization in paramagnetic materials, and does not directly rely on the presence of a phase transition.

Adiabatic demagnetization was first investigated more than a century ago and was proposed as a cooling method not long after \cite{Smith2013-MCEHistory}. The technique requires a material with a large magnetic susceptibility that remains paramagnetic to low temperatures. In 1933, adiabatic demagnetization was successfully used for the first time at cryogenic temperatures to cool Gd$_2$(SO$_4$)$_3\cdot$8H$_2$O from 1.5 K to 0.25 K by cycling a magnetic field of 0.8 T \cite{Giauque1933-ADR}. Since then, there have been many advances in the materials used for magnetic refrigeration, both at cryogenic temperatures and near room temperature \cite{Kamran2020-MagRefReview,Li2020-MCEreview,Balli2017-MCEreview,Frano2018-MCEreview,Kitanovski2020-MCEreview,Gschneidner2008-MCEreview,Moya2014-caloricReview}.

Low temperature Jahn-Teller systems are especially appropriate for cryogenic cooling via an analogous adiabatic decompression process for two reasons. Firstly, they preserve a large amount of entropy to low temperature, owing to the degeneracy of CEF levels. Secondly, they have a large magnetoelastic coupling, which allows for the ground state, and thus the entropy, to be easily tuned with strain. Our focus of this study is not to propose a particular material for which low temperature cooling is optimized, but rather to experimentally verify the entropy landscape of an archetypical material in this class, TmVO$_4$. In so doing, we also demonstrate an especially suitable and effective method to measure a quantity of fundamental interest in the study of quadrupole order, the dynamic quadrupole strain susceptibility \cite{Rosenberg2019-Diverging,Ikeda2021-ECEB2gFluc}

TmVO$_4$ undergoes a cooperative Jahn-Teller phase transition at $T_Q = 2.15$ K to a ferroquadrupolar ordered state. The phase transition is continuous, and the crystal symmetry changes from tetragonal to orthorhombic \cite{Cooke1972-ortho}, corresponding to a spontaneous $\varepsilon_{xy}$ shear strain \cite{Segmuller-SponStrain}. As we will show, in the para-quadrupolar regime (i.e. for temperatures above the phase transition), the material has a large quadrupole-strain susceptibility with respect to $xy$ strains, much as a ferromagnetic material has a large magnetic susceptibility above its Curie temperature. In terms of entropy, above $T_Q$, each Tm ion contributes $k_B \ln 2$ due to the crystal field groundstate, which is a non-Kramers doublet \cite{Knoll1971-CEFspectra}. This entropy is removed at the phase transition, which splits the ground state doublet \cite{Segmuller-SponStrain}. To the extent that relevant terms in the Hamiltonian are known \cite{Gehring1975-review}, it is possible to estimate the entropy as a function of temperature and strain (Fig. 1). Of particular significance, the large quadrupole strain susceptibility means that the entropy can be substantially reduced above $T_Q$ by application of modest shear strains (blue curve in Fig. 1). Such large changes in entropy at low temperatures in response to modest applied strains implies the possibility of a sequence of isothermal compression and adiabatic decompression (black arrows in Fig. 1) to achieve a substantial cooling effect for temperatures above $T_Q$. Such large temperature changes ($\Delta T/T \sim 1$ under adiabatic conditions, or equivalently $\Delta S/S \sim 1$ under isothermal conditions) justifies the moniker `giant' for the elastocaloric effect in this material.

\begin{figure}[t]
\centering
  \includegraphics[width=\linewidth]{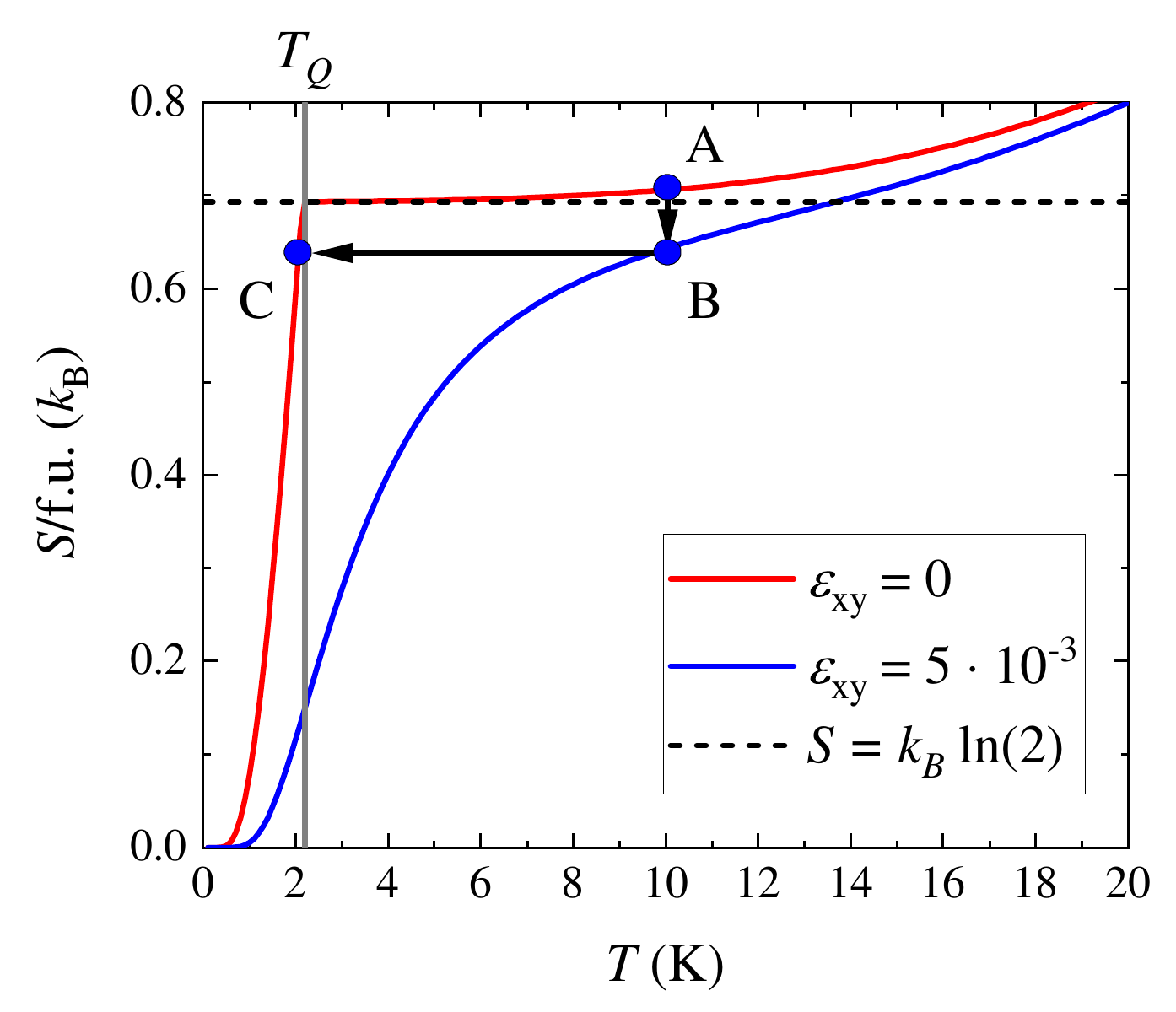}
  \caption{
\textbf{The calculated entropy of TmVO$_4$ as a function of temperature for $\boldsymbol{\varepsilon_{xy} = 0}$ and $\boldsymbol{5\cdot 10^{-3}}$, expressed per formula unit (f.u.).} The $4f$ contribution is calculated from the Hamiltonian using coefficients from published literature while the phonon contribution is determined from heat capacity experiments (Supplementary Fig. S11).  In the absence of externally-induced strain ($\varepsilon_{xy} = 0$), the entropy falls rapidly from approximately $k_B \ln{2}$ above $T_Q$ towards zero at zero temperature. Inducing a modest $xy$ strain of just $5\cdot 10^{-3}$ leads to a substantial reduction of the entropy for all temperatures below approximately 15 K (see main text), while also removing the phase transition. An example of a possible thermodynamic cycle is indicated (black arrows). The cycle consists of an isothermal path along which $\varepsilon_{xy}$ is increased ($A\rightarrow B$) and an adiabatic path along which $\varepsilon_{xy}$ is decreased ($B\rightarrow C$). A thermalization step can be included from $C\rightarrow A$ to produce a full cycle. The figure demonstrates that a large cooling effect can, at least in principle, be realized if adiabatic conditions are maintained during the decompression part of the cycle.}
\end{figure}

\begin{figure*}[t]
\centering
  \includegraphics[width=\textwidth]{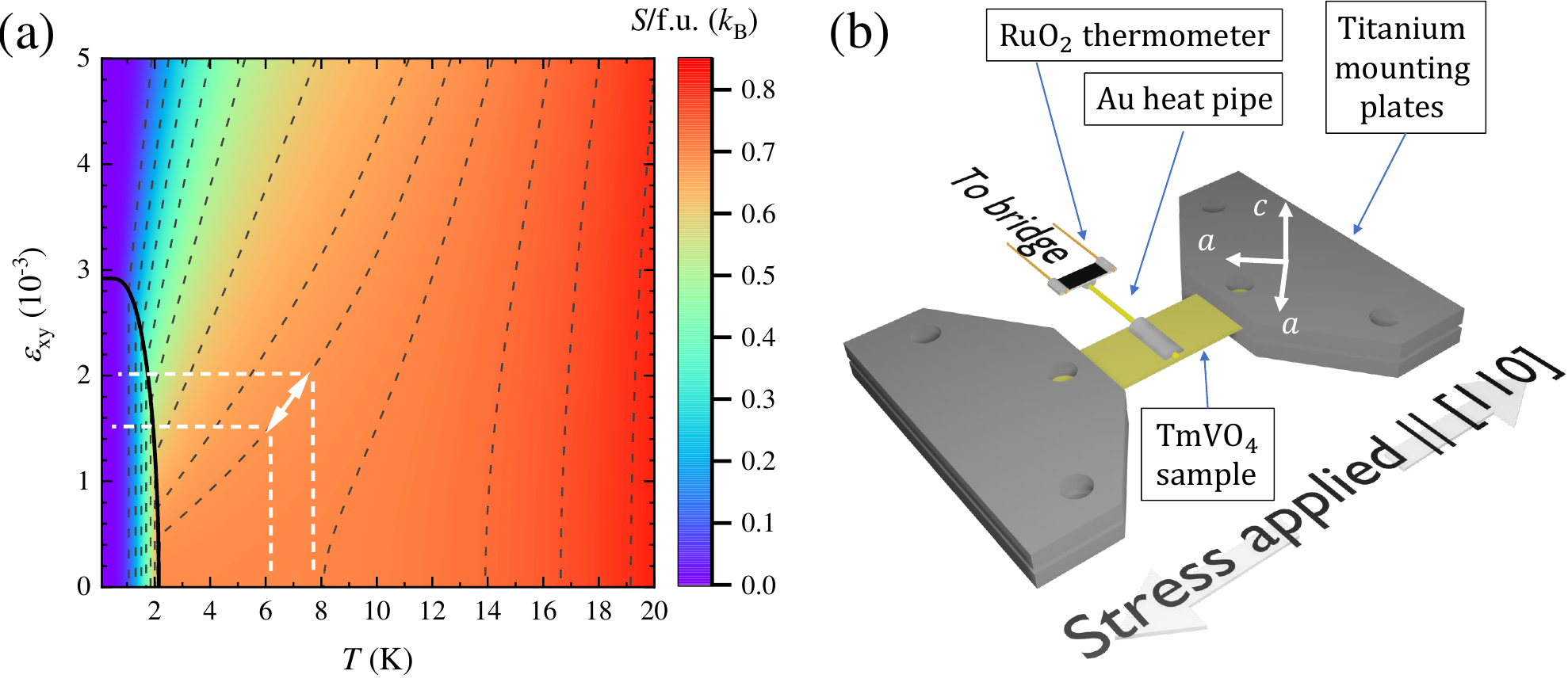}
\caption{
\textbf{Calculated entropy of TmVO$_4$ and experimental ECE sample mounting setup.} (a) The calculated entropy landscape of TmVO$_4$ as a function of $T$ and $\varepsilon_{xy}$, shown here as a heat map. Three broad regions can be identified. At high temperatures, phonons dominate the entropy and isentropic curves are nearly vertical; cooling potential is low in this region. At low temperatures and larger strains, the isentropic curves become almost radial lines (emanating from the origin), as the $4f$ contribution to the entropy outweighs the phonon contribution; potential for cooling by adiabatic decompression is optimized in this region. Finally, below $T_Q=2.15$ K, a first order transition separates a region in which the material is homogeneously strained (above and to the right of the solid black line), from a region in which domains of oppositely oriented orthorhombic regions are energetically favored (below and to the left of the solid black line). Neglecting the small contribution to the total entropy that arises from the domain walls, the total entropy does not exhibit a strain-dependence within this region below $T_Q$, providing a practical limit to the lowest possible temperature that could be achieved via adiabatic cooling for realistic strains.  White arrow (exaggerated for clarity) indicates the AC ECE method that we employ to experimentally determine the entropy landscape (see main text). (b) Schematic diagram illustrating our low-temperature AC ECE experimental setup. A single crystal of TmVO$_4$ is held between two titanium mounting plates. The crystal is cut such that the [110] direction lies along the stress direction. We apply both AC and DC voltages across different piezoelectric stacks (not shown) to control the distance between the titanium mounting plates, thus controlling the strain the material experiences. The strain oscillations produce temperature oscillations that are measured by a RuO$_2$ thermometer (see Supplementary Materials for details). Typical sample dimensions are of order 1.6 mm in length (of which approximately 0.9 mm ‘active area’ extends between the titanium plates) by 0.3 mm width by 0.05 mm thickness.
}
\end{figure*}

The primary goal of the present work is to experimentally verify the anticipated entropy landscape and demonstrate the presence of the associated giant elastocaloric effect at low temperatures in TmVO$_4$. Measured using an AC (dynamic) elastocaloric technique \cite{Ikeda2019-ECEtechnique, Ikeda2021-ECEB2gFluc,Li2022-MackenzieECE}, the AC analog of the process one would employ for a viable cooling technology, our observations also provide a direct measure of the dynamic quadrupole strain susceptibility.

The measurement can be best understood by considering the estimated entropy landscape (Fig. 2(a)). Under adiabatic conditions, appropriate for sufficiently rapid strain osillations, the material moves back and forth along a constant entropy contour in response to an externally-induced AC strain, resulting in a concomitant temperature oscillation at the same frequency (white arrow in Fig. 2(a)). As temperature is lowered towards the ferroquadrupolar phase transition, the constant entropy curves decrease in slope, yielding a larger amplitude temperature oscillation for a given amplitude strain oscillation. This is the regime in which the entropy is dominated by the $4f$ electrons, in contrast to the higher temperature regime (above approximately $15$ K), where the phonons, which have a much smaller strain-dependence, dominate the total entropy. The cooling effect in this material is thus primarily restricted to below approximately $15$ K. 

We measure the elastocaloric response of the material and map out the entropy in temperature-strain space by cooling the sample and inducing an $xy$ strain in the TmVO$_4$ single crystal via stress applied along the tetragonal [110] direction (Fig. 2(b)). To apply this stress, we use a commercially available Razorbill CS100 cell. The sample is mounted with the tetragonal [110] direction aligned along the displacement direction and we measure the temperature response of the sample with respect to an induced AC strain using a RuO$_2$ temperature sensor connected to a Wheatstone bridge (Supplementary Fig. S1). Simultaneously, we induce an offset strain and change the temperature of the sample to navigate across the $T$-$\varepsilon_{xy}$ entropy landscape (Fig. 2(a)).

\begin{figure}[t]
\centering

  \includegraphics[width=\linewidth]{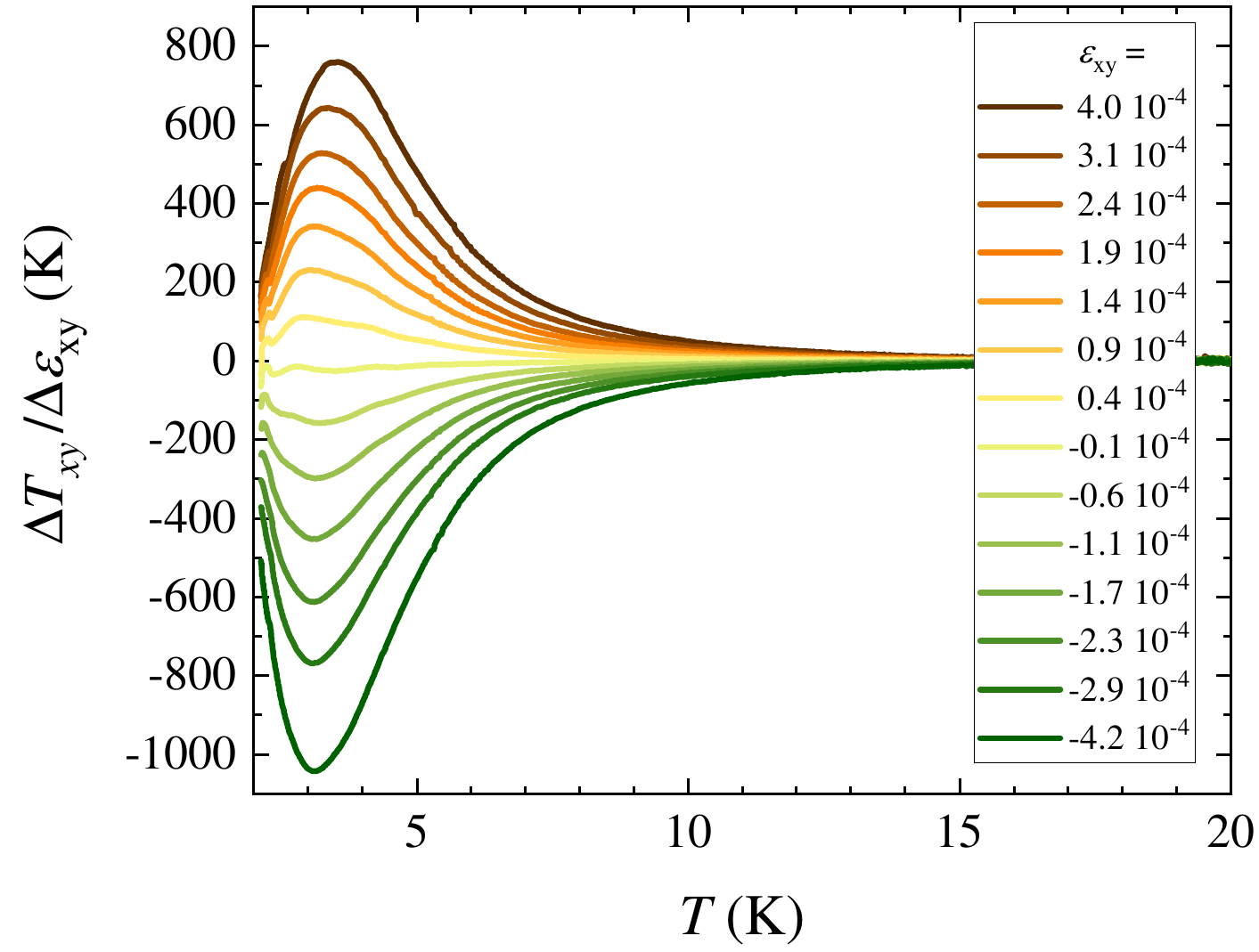}
  \caption{
  \textbf{Temperature-dependence of the ECE ($\boldsymbol{\Delta T_{xy}/\Delta \varepsilon_{xy}}$) for several representative offset strains.}  Data were taken by applying a fixed voltage to the piezoelectric stacks in the strain cell and then sweeping temperature, and are labeled with the strain the sample experiences at 10 K, which changes slightly over the entire temperature range (Supplementary Fig. S6). The data reveal a characteristic ‘fish tail’ shape, discussed in the main text.
}
\end{figure}

\begin{figure}[t]
\centering
  \includegraphics[width=\linewidth]{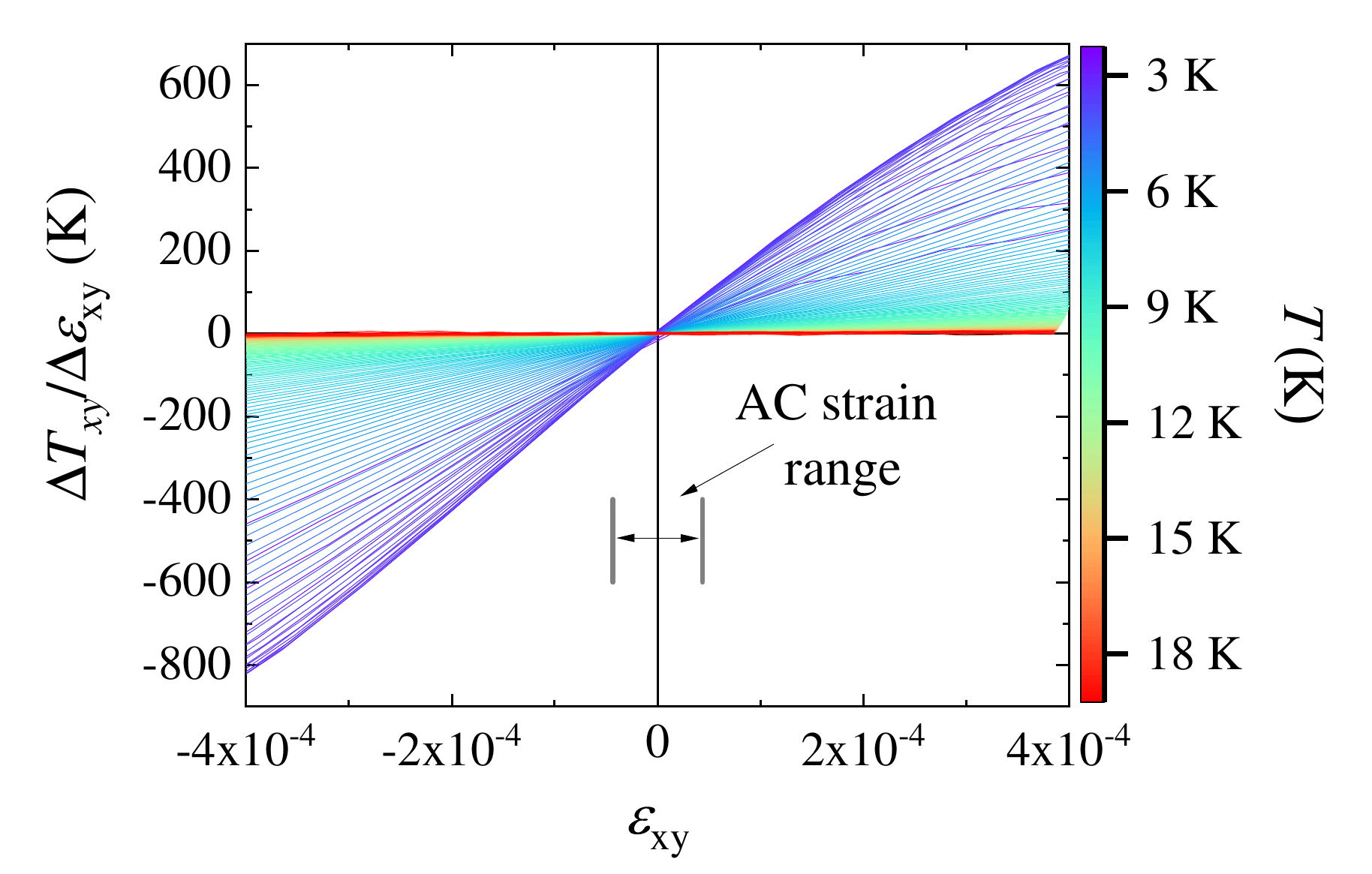}
  \caption{\textbf{Strain dependence of the ECE for different temperatures.} The amplitude of the AC strain used for the measurements is indicated for comparison. The linear dependence of the ECE for small values of $\varepsilon_{xy}$ is apparent.
  }
\end{figure}

\begin{figure}[t]
\centering
  \includegraphics[width=\linewidth]{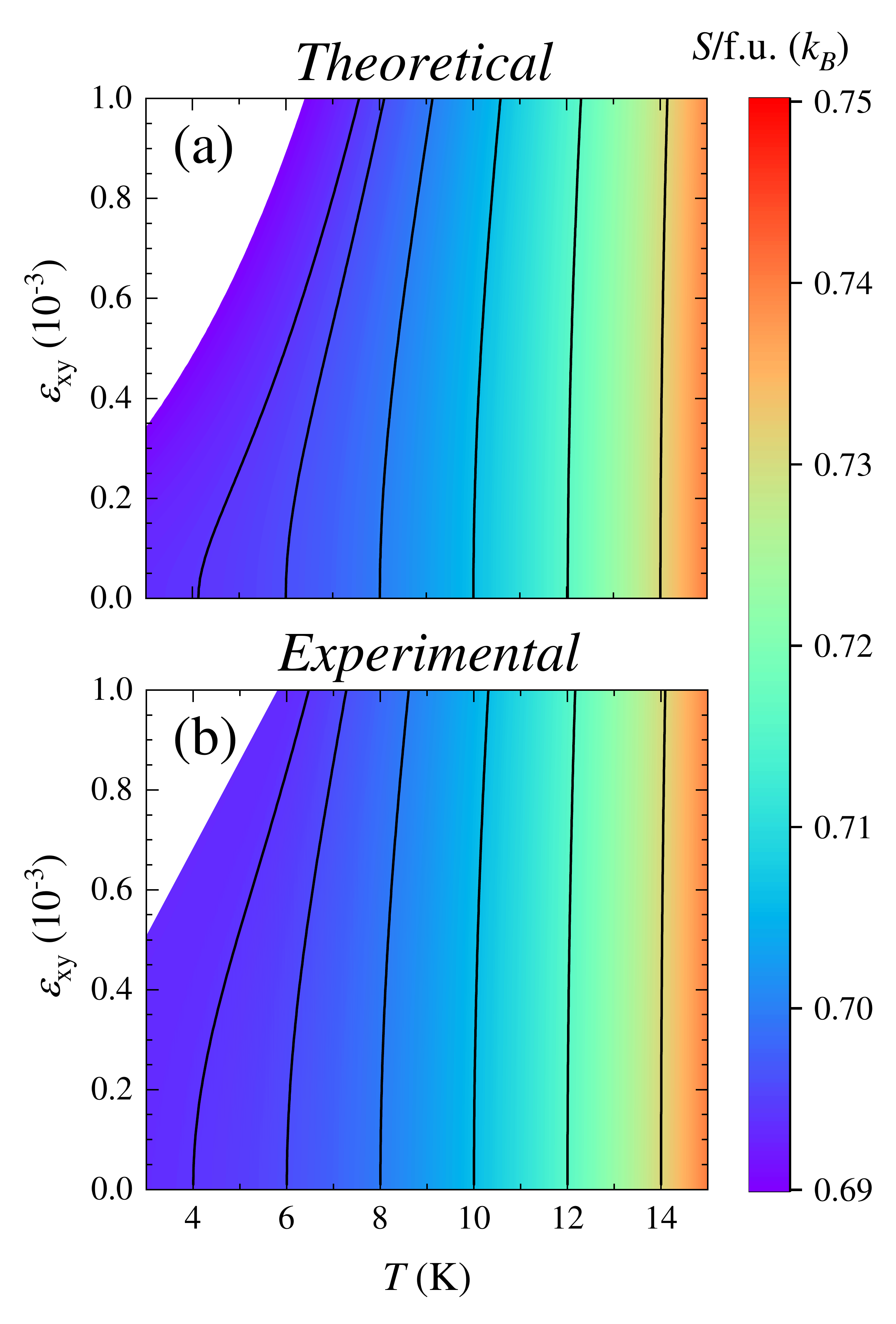}
\caption{\textbf{A comparison of the entropy landscape (a) calculated from the Hamiltonian to (b) those obtained from ECE measurements, after scaling to ultrasound measurements.} Both figures are plotted with constant entropy contours to emphasize similarity. The experimentally realized entropy landscape matches well with theoretical predictions.}
\end{figure}

\section*{Results}

\subsection*{Reconstructing the entropy landscape}

We adopt a coordinate system in which un-primed coordinates refer to the principle crystal axes, and primed coordinates are rotated by 45 degrees about the c-axis.

Stress applied along the [110] crystallographic direction induces a combination of symmetric and antisymmetric (i.e. shear) strains. Using a reference frame in which x' is along the stress direction, the total observed elastocaloric effect, $\Delta T/\Delta \varepsilon_{x'x'}$ reflects contributions from both the symmetric and antisymmetric strains. Since the elastic stiffness tensor of TmVO$_4$ is known \cite{Melcher1973-elastic-constants}, measurement of $\varepsilon_{x'x'}$ under conditions of uniaxial stress is sufficient to determine all of the terms in the elastic stiffness tensor, and simple coordinate transformations can be used to go between un-primed and primed coordinate systems. Details associated with strain relaxation effects are discussed in the Supplementary Material.

Of the two symmetry-decomposed strains that the sample experiences as a consequence of the [110] stress, antisymmetric strain has a much larger effect on the entropy, and hence on the measured elastocaloric temperature oscillations. Shear strain, $\varepsilon_{xy}$, couples to the quadrupole moments bilinearly (i.e. $H_{eff} \sim \varepsilon_{xy} P_{xy}$, where $P_{xy}$ is the operator that describes the $xy$ quadrupole, $P_{xy}=1/2(J_xJ_y+J_yJ_x)$, and hence has a large effect on the entropy at low temperatures. In contrast, symmetric strain (both $1/2(\varepsilon_{x'x'} + \varepsilon_{y'y'})$ and $\varepsilon_{zz}$) does not split the CEF groundstate above $T_Q$, and consequently has a much smaller effect on the observed total elastocaloric effect. The small contribution to the elastocaloric temperature oscillations that arises from symmetric strains can be readily identified based on symmetry grounds (see Supplementary Material), and subtracted from the total signal to yield the part of the ECE that arises purely from $xy$ shear: $\partial T/ \partial \varepsilon_{xy}$. This quantity is plotted as a function of temperature for several representative DC offset strains in Fig. 3. These data were taken by sweeping temperature while a fixed bias voltage was applied to the outer stacks of the CS100 strain cell. Strain values are listed at 10 K; subtle changes in strain with temperature are accounted for in the subsequent analysis and are used here only as labels.  

Inspection of Fig. 3 reveals several key features. First, the observed ECE is odd with respect to the bias strain. That is, the sign of the ECE is either positive or negative depending on the sign of the bias strain. This is a natural consequence of antisymmetric deformations for which the AC component either adds or subtracts from the DC bias strain. Second, the data reveal a characteristic ‘fishtail’ shape, in which the maximum value is observed close to 3 K, reflecting the slope of the entropy contours indicated in Fig. 2. Finally, the magnitude of the temperature oscillations is seen to be very large. For example, for a bias strain of only $4 \cdot 10^{-4}$ (0.04 \%), the amplitude of the temperature oscillation (normalized by the amplitude of the AC strain oscillation, which is $5 \cdot 10^{-5}$) is approximately 700 K in dimensionless units of strain, or equivalently 7 K per percent of strain. 

The data shown in Fig. 3 can be replotted, now as a function of strain, $\varepsilon_{xy}$; this is shown in Fig. 4. Because the strain is recorded for every temperature, these curves do not suffer from the same concerns about strain changing as a function of temperature. Data are shown for a range of temperatures, each a different color. For comparison, the AC strain amplitude is also indicated. As can be readily seen, for small strains, $\partial T/ \partial \varepsilon_{xy}$ varies linearly with $\varepsilon_{xy}$, and indeed, as mentioned above, is an odd function of $\varepsilon_{xy}$.

Despite the large values of the ECE that we observe, these values are uniformly approximately a factor of five smaller than theoretical values due to imperfect adiabaticity of the experiment. Finite element simulations confirm the imperfect adiabaticity, which can be corrected for by multiplying by a constant, temperature-independent, factor (Supplementary Fig. S15). Doing so does not change any of the subsequent conclusions, including the temperature- and strain-dependence of the entropy, and the temperature dependence of the quadrupole strain susceptibility.

The above data can be used to reconstruct the entropy landscape. Starting from the measured heat capacity as a function of temperature, simple integration with respect to temperature yields $S(T, \varepsilon_{xy} = 0)$. Because the experiment measures $\partial T/ \partial \varepsilon_{xy}$  for adiabatic conditions, a process of numerical iteration can be used to sequentially build up the entire entropy landscape. The result of performing this process is shown in Fig. 5 with the entropy shown as a colorscale. Here we have included the factor of five mentioned above in order to capture the full entropy. This entropy analysis can be compared with the anticipated theoretical result, using the known Hamiltonian and magnetoelastic coupling coefficients \cite{Melcher1976-Review}. Starting from the measured zero-strain heat capacity, the obtained entropy landscape (Fig 5(b)) closely matches that determined from the Hamiltonian (Fig 5(a)).

\subsection*{Determining the quadrupole strain susceptibility}

Further analysis of the data presented above yields an estimate of the quadrupole strain susceptibility. To understand this, it is helpful to briefly consider the thermodynamic quantity that is being measured. Specifically, 
\begin{equation}
    \frac{\partial T}{\partial \varepsilon_{xy}} = -\bigg(\frac{\partial S}{\partial \varepsilon_{xy}}\bigg)_T/\bigg(\frac{\partial S}{\partial T}\bigg)_{\varepsilon_{xy}} = \frac{\partial c_{66}}{\partial T}\varepsilon_{{xy}} \frac{T}{C_{\varepsilon}}
\end{equation}
where we have used a Maxwell relation to relate the strain derivative of the entropy to the temperature derivative of the elastic modulus $c_{66}$, and $C_\varepsilon$ is the heat capacity under constant strain. We use the volumetric heat capacity in our subsequent analysis to allow for a direct comparison with ultrasound measurements. For materials that undergo a continuous, $xy$-symmetric quadrupolar (or more generally nematic\footnote{An electronic nematic state breaks the discrete rotational symmetry of the crystal lattice. The ferroquadrupolar order discussed in the present work is a specific realization of such an electronic nematic state. See for example Maharaj, \emph{et al.}\cite{Maharaj2017-TFIM}}) phase transition, the temperature dependence of $c_{66}$ is given, quite generally, by $c_{66}(T) =  c_{66}^0 - \lambda^2 \chi_{B_{2g}}$, where $c_{66}^0$ is a ‘bare’ (unrenormalized) elastic stiffness, $\lambda$ is the coupling constant that relates strain and the order parameter (i.e. the free energy contains a term $\Delta F \sim \lambda \varepsilon_{B_{2g}} \Psi_{B_{2g}}$), and $\chi_{B_{2g}} = \partial\braket{P_{xy}}/\partial\varepsilon_{xy}$ is the quadrupole strain susceptibility \cite{Ikeda2021-ECEB2gFluc}. In other words, the measured adiabatic elastocaloric coefficient is proportional to the temperature derivative of the quadrupole strain susceptibility. And because this is measured at a non-zero frequency, it is technically a dynamic susceptibility (albeit, in this case, strictly in the DC limit given the measurement frequency of 211 Hz). All that is required to obtain this quantity is to extract the slope of $\partial T/\partial \varepsilon_{xy}$ with respect to $\varepsilon_{xy}$ (i.e. the slope of the data shown in Fig. 4) and multiply by the measured heat capacity. In the absence of heat capacity measurements performed \emph{in situ}, we use data obtained from a standard relaxation technique for which subtle differences in strain conditions can affect the heat capacity (Supplemental Fig. S11) and consequently the extracted susceptibility at the lowest temperatures. To compensate for strain effects, we have corrected the heat capacity to better match the low-temperature ultrasound data (Supplemental Fig. S13). There are no differences above 6 K and the low temperature upturn remains a robust feature regardless of subtle differences in the heat capacity.

Fig. 6 shows the result of performing this analysis, with details described in the Supplementary Material. As can be seen, the data diverge towards low temperature. More specifically, the data can be very well fit to $A(T-\theta)^2$, the anticipated functional form assuming $\chi_{B_{2g}}$ follows Curie-Weiss behavior. It is precisely this diverging quadrupole susceptibility that permits the large elastocaloric effect, and supports our proposal to use adiabatic decompression of such materials for cryogenic cooling.

\begin{figure}[t]
\centering
  \includegraphics[width=\linewidth]{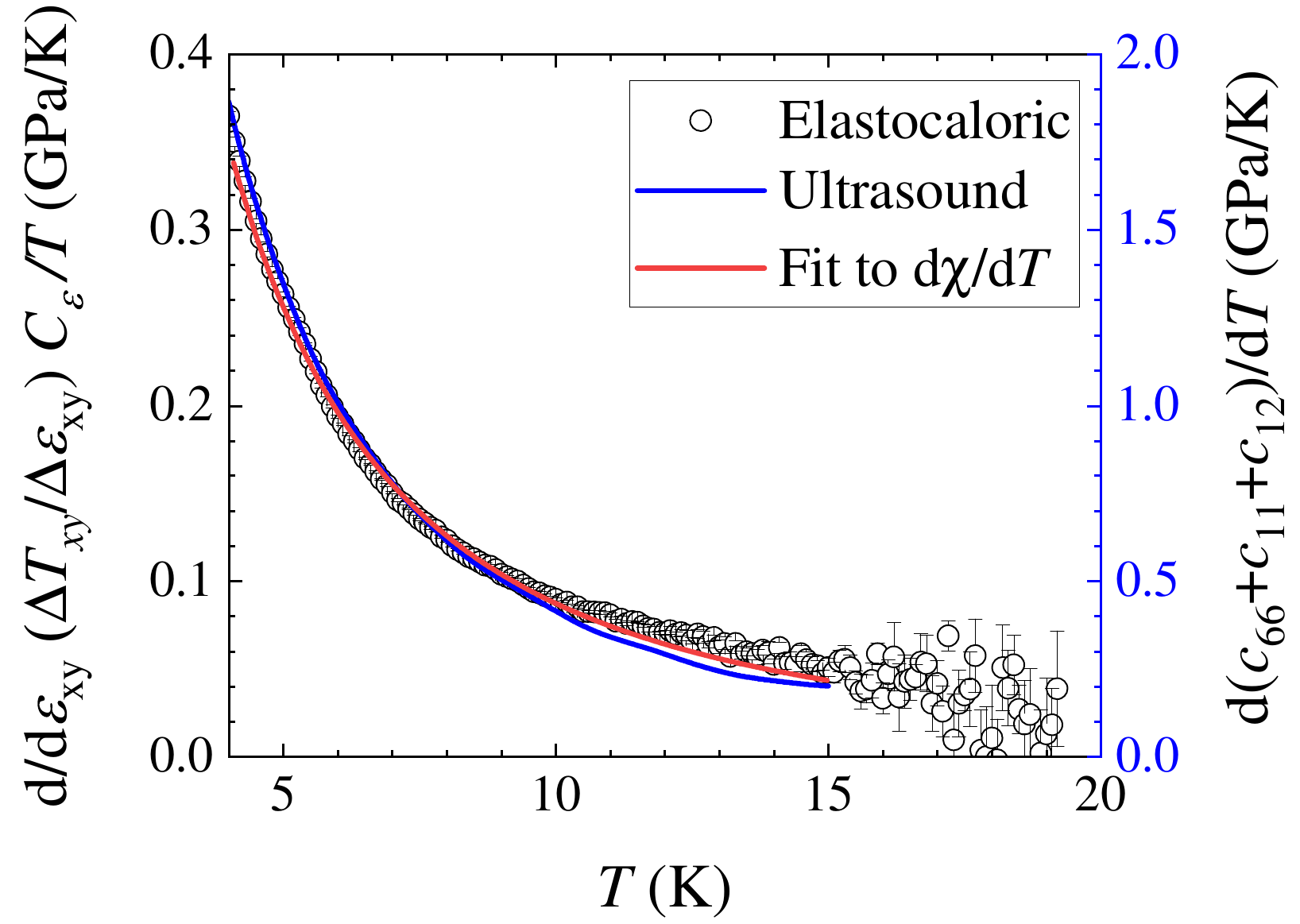}
  \caption{\textbf{Divergence of the quadrupole strain susceptibility of TmVO$_4$.} Red line shows a fit to the anticipated Curie-Weiss functional form, with fit parameters matching previously determined values (see main text). Right hand axis (blue data) shows the temperature derivative of the elastic moduli ($c_{66}+c_{11}+c_{12}$), obtained from ultrasound measurements (see main text). Agreement between the functional form of the two measurements indicates that imperfect adiabaticity in the ECE measurement does not contribute any additional temperature-dependence to the measured signal.
  }
\end{figure}

Because of the relationship between the ECE and $\partial c_{66}/\partial T$, we can also compare our measurement to direct measurements of the elastic stiffness tensor, in this instance from pulse-echo ultrasound measurements. For a carefully chosen experimental geometry (see Methods and Supplemental Material), the speed of sound obtained from these pulse-echo measurements yields an estimate for the specific combination of stiffness coefficients  ($c_{66} + c_{11} + c_{12}$). Since the singular behavior associated with $\chi_{B_{2g}}$ only affects $c_{66}$, any weak temperature dependence associated with $c_{11}$ and $c_{12}$ is overwhelmed by that of $c_{66}$, and thus taking the temperature derivative of this quantity yields a result that is predominantly determined by $\partial \chi_{B_{2g}}/\partial T$. The agreement between the ECE and pulse-echo measurements (and earlier estimates of $c_{66}$\cite{Melcher1973-elastic-constants}) is remarkable, as is the agreement with the anticipated Curie-Weiss functional form.

\section*{Discussion}

We have demonstrated that TmVO$_4$ displays a giant elastocaloric effect at low temperatures (Fig. 4). The large magnitude reflects a large (diverging) quadrupole strain susceptibility (Fig. 6), and ultimately derives from a combination of a large magnetoelastic coupling and the preservation of a large entropy to low temperatures, both of which are implicit in such a Jahn-Teller active material for which the cooperative phase transition occurs at such a low temperature. The material is not unique in this regard, and other examples can be readily imagined for which a similarly large, low-temperature ECE can be anticipated. Indeed, a search for similar materials with even larger magnetoelastic coefficients could result in materials for which strain is an even more effective tuning parameter. 

The principle of cooling via adiabatic decompression of para-quadrupolar (or more generally para-nematic) materials, as we have outlined here, is straightforward. Such a method has several distinct advantages over traditional methods for cooling in the temperature regime below 2 K, for which either adiabatic demagnetization or $^3$He (either pumped $^3$He or as a mixture with $^4$He in a dilution refrigerator) are standard choices. First, this is a ‘dry’ technology, capable, at least in principle, of taking a 10 K sample stage down to sub Kelvin base temperatures without specifically requiring the use of liquid helium, and thus reducing the need for complex gas/liquid handling systems. In contrast to adiabatic demagnetization, elastocaloric cooling does not require application of magnetic field, which saves on costly infrastructure and can be used in applications for which zero magnetic field is a requirement. In addition, the piezoelectric actuators that we have used to control the strain in these experiments have rapid response times (much less than a millisecond), in contrast to ramping a magnetic field for adiabatic demagnetization (which takes several minutes), which in principle enables rapid cycling of the thermal cycle, therefore increasing cooling power. Finally, in contrast to typical higher-temperature ECE cooling technologies, adiabatic decompression of para-quadrupolar materials specifically uses a working material that is always kept in a homogeneous state above its critical temperature, significantly reducing concerns over sample fatigue.

\subsection*{Methods}

TmVO$_4$ crystals were grown using a molten flux technique by dissolving Tm$_2$O$_3$ powder into Pb$_2$V$_2$O$_7$ flux as previously described \cite{Feigelson1968-Growth,Smith1974-Growth}. The sample used for the experiments performed here was cut from a large sample such that the long axis, and hence the stress applied, is along [110]. For the measurements presented, the sample was affixed to a Razorbill CS-100 cell using 5-minute and 2-ton epoxy. The dimensions of the strained portion of the sample measured approximately 950 $\mu$m $\times$ 360 $\mu$m $\times$ 65 $\mu$m ($l \times w \times t$). The sample was coated with a very thin layer of Angstrom Bond to reduce the likelihood of the sample breaking. Temperature oscillations were measured using a ruthenium oxide Wheatstone bridge, with one thermometer mounted via a gold wire heat pipe on the sample with DuPont 4929N silver paste (to eliminate the elastoresistance response of the thermometer) and the others mounted nearby but off of the sample. A rendering of the experimental setup is shown in Fig. 2(b). We induce DC offset strains simultaneously with our probing AC strain via piezoelectric stacks. More details on the thermometry setup are included in the Supplementary.

Heat capacity measurements were completed using a Quantum Design Physical Properties Measurement System using the standard 2$\tau$ relaxation technique. In these measurements, the sample is affixed to a platform using Apiezon N-grease, which freezes, and therefore stresses and strains the crystal via differential thermal contraction \cite{Massat2022-PNAS} (see Supplementary for more details). 

The temperature dependence of $c_{66}$ was measured using a standard phase-comparison pulse echo technique. Single-crystal samples of TmVO$_4$ were prepared with two parallel faces normal to the $[110]$ and $[\Bar{1}\Bar{1}0]$ directions. The faces were aligned to better than 1 degree using Laue diffraction and polished using diamond lapping film. Thin-film, ZnO, longitudinal-mode transducers were RF sputtered using a ZnO target in a mixed oxygen/argon atmosphere. 

Measurements were performed in an Oxford Instruments Heliox $^3$He refrigerator. Short bursts (typically $\sim$ 50 ns) of radiofrequency signals, with the carrier frequency of approximately 500 MHz, were generated with a Tektronix TSG 4106A RF generator modulated by a Tektronix AFG 31052 arbitrary function generator, amplified by a Mini-Circuits ZHL-42W+ power amplifier, and transmitted to the transducer. The signal was detected with the same transducer, amplified with a Mini-Circuits ZX60-3018G-S+ amplifier, and recorded on a Tektronix MSO64 oscilloscope. The detection amplifier was isolated from the power amplifier using Mini-Circuits ZFSWA2-63DR+ switches, timed with the same Tektronix AFG 31052 arbitrary function generator.

\subsection*{Acknowledgments}
This work, including ECE measurements and analysis, were supported by the Air Force Office of Scientific Research under award number FA9550-20-1-0252. MPZ was also partially supported by a National Science Foundation Graduate Research Fellowship under grant number DGE-1656518. Initial crystal growth (PM) and ECE technique development (MSI and EWR) was supported by the Gordon and Betty Moore Foundation Emergent Phenomena in Quantum Systems Initiative through Grant GBMF9068. Entropy analysis (LY) was supported by the National Science Foundation (DMR- 2232515). Ultrasound measurements (PMH and BJR) were supported by the Office of Basic Energy Sciences of the United States Department of Energy under award no. DE-SC0020143.

\bibliography{TmVO4-B2g-ECE-NatMater}

\end{document}